# Strong quantum confinement effect in the optical properties of ultrathin α-In$_2$Se$_3$

*Jorge Quereda[1,*], Robert Biele[2], Gabino Rubio-Bollinger[1,3,4], Nicolás Agraït[1,5], Roberto D'Agosta[2,6], and Andres Castellanos-Gomez[5,*]*

[1]*Departamento de Física de la Materia Condensada. Universidad Autónoma de Madrid, Madrid, E-28049, Spain.*
[2]*Departamento de Física de Materiales, Universidad del Pais Vasco, EHU/UPV, San Sebastian E-20018, Spain.*
[3]*Instituto de Ciencia de Materiales Nicolás Cabrera, E-28049, Madrid, Spain.*
[4]*Condensed Matter Physics Center (IFIMAC), Universidad Autónoma de Madrid, E-28049, Madrid, Spain.*
[5]*Instituto Madrileño de Estudios Avanzados en Nanociencia (IMDEA – Nanociencia), E-28049, Madrid, Spain.*
[6] *IKERBASQUE, Basque Foundation for Science, Bilbao E-48013, Spain.*
Jorge.quereda@uam.es , andres.castellanos@imdea.org

The isolation of atomically thin semiconductors has attracted the interest of the nanoscientists as these materials can complement graphene in those applications where its lack of an electronic band gap hinders its use.[1–5] In fact, two dimensional semiconductors have been recently employed in the fabrication of field effect transistors, [6,7] photodetectors [8–14] and solar cells.[15–19] Inherent to their reduced out-of-plane dimension, atomically thin semiconductors present a marked thickness-dependent band structure due to the quantum confinement of the charge carriers. For instance, the band gap in Mo- and W- based transition metal dichalcogenides increases from its bulk value (~1.3 eV, indirect gap) up to 1.9 eV (direct gap) in the single-layer limit.[20,21] More recently, black phosphorus has also shown a pronounced thickness-dependent band gap ranging from ~0.3 eV (direct gap) for bulk to ~1.75 eV (direct gap) for single-layer black phosphorus.[22] This thickness-dependent band gap can be very advantageous for applications as photodetectors as one can select the sensing photon energy window by simply selecting the right material thickness for the photodetector device. Up to now, however, there is a broad spectral window from 2.0 eV to 3.0 eV uncovered by 2D materials that could be very relevant for applications requiring 'solar blind' semiconducting materials (not absorbing within the visible spectrum) but responsive to near-UV light.

In this work we demonstrate a very strong quantum confinement effect in the optical properties of atomically thin α-In$_2$Se$_3$ crystals. We observe a marked thickness-dependent shift in the optical absorption spectra acquired on mechanically exfoliated In$_2$Se$_3$ flakes with thickness ranging from 3.1 nm (~ 3 layers) up to 95 nm (more than 90 layers). While for the thicker flakes we extract band gap values of 1.45 eV (in good agreement with the experimental[23] and theoretical values of bulk α-In$_2$Se$_3$), thin flakes show a remarkable increase of their optical band gap, reaching 2.8 eV for the thinnest flakes studied here. To better understand and support our experimental findings, we have performed density functional theory (DFT) calculations in combination with many-body techniques. Our calculations serve to obtain the band structure for bulk α-In$_2$Se$_3$ from where we extract the effective mass of the exciton and the electronic band gap of 1.4 eV in good agreement with the experiment. Furthermore, we have also calculated the optical gap of 1.32 eV by solving the Bethe-Salpeter equation (BSE) indicating a relatively small binding energy of the exciton. Details of the calculation can be found in the Supporting Information.

**Figure 1**a shows an optical image of an In$_2$Se$_3$ crystal with regions of different thickness, as well as an AFM scan profile along the dashed line, fabricated by micromechanical exfoliation as described in the experimental section. An AFM image of the same flake from Figure 1 is shown in the Supporting Information (Figure S1). In order to study the optical absorption of the In$_2$Se$_3$ crystals we used a home-made hyperspectral imaging setup, described in detail elsewhere.[25] The hyperspectral imaging is carried out by sweeping the excitation wavelength in steps and acquiring transmission mode images of the In$_2$Se$_3$ crystals for each wavelength. The collected data is then arranged in a three-dimensional matrix, being the first two matrix

indexes the *X* and *Y* spatial coordinates and the third index the wavelength, *λ*. Spectral information of a certain sample region can be directly obtained by plotting all the elements along the wavelength dimension *λ* for given *X* and *Y* coordinates.

Figure 1b shows the wavelength-dependent optical transmittance at five In$_2$Se$_3$ regions of different thicknesses, from 3.1 nm to 25.1 nm. Interestingly, we find that for any given illumination wavelength the optical transmittance decreases monotonically with the flake thickness, as shown in the inset of Figure 1b for *λ* = 550 nm. Therefore, one can estimate the thickness of In$_2$Se$_3$ by quantitatively measuring the transmittance at a fixed illumination wavelength. This also enables us to generate a thickness map from the acquired hyperspectral images, see Figure 1c. The inset in Figure 1c shows a line profile along the same region measured in the AFM scan of Figure 1a. The obtained thickness is in good agreement with that measured by AFM. We address the reader to the Supporting Information for details about the analysis employed to generate the 2D thickness maps from the hyperspectral datasets.

The absorption coefficient (*α*) of In$_2$Se$_3$ can be calculated from its optical transmittance (*T*) and its thickness (*d*) as

$$\alpha(\lambda) = -\frac{\log_{10}(T(\lambda))}{d} \ . \tag{1}$$

According to Tauc *et al.*[26], near the absorption edge the absorption coefficient of a direct-gap semiconductors is

$$\alpha = A\frac{\left(\hbar\omega - E_g^{\text{opt}}\right)^{1/2}}{\hbar\omega} \ , \tag{2}$$

where *A* is a material-dependent constant and $E_g^{\text{opt}}$ is the optical energy gap of the material. Equation (2) can be rewritten as

$$(\alpha\hbar\omega)^2 = A^2(\hbar\omega - E_g^{\text{opt}}) \ . \tag{3}$$

The usual method for determining the value of $E_g^{opt}$ involves representing $(\alpha\hbar\omega)^2$ versus the photon energy, $\hbar\omega$, and fitting the absorption band edge to a linear function. According to Equation (3), the intersection of this linear fit with the horizontal axis yields the value of the optical bandgap. **Figure 2**a shows the Tauc plots for seven $In_2Se_3$ flakes with different thicknesses, from 3.1 nm to 28.6 nm, as well as their linear fit near the absorption edge. We find that the optical bandgap is strongly dependent on the flake thickness, increasing from its bulk value of ~1.45 eV to a value of 2.8 eV for the thinnest measured flakes.

It is also possible to spatially map the variations of the $In_2Se_3$ band gap from the acquired hyperspectral data as for each set of *X, Y* coordinates of the hyperspectral data matrix one can extract an absorption spectra. The inset in Figure 2b shows a colormap representing the energy at which the absorption reaches a cutoff value of $10^5$ cm$^{-1}$ at each sample position (isoabsorption map). Note that when a small absorption cutoff is used to construct the isoabsorption map, the isoabsorption energy resembles the optical band gap energy and thus the isoabsorption map represents accurately the spatial variation of the band gap. Figure 2b shows the band gap energies extracted from the isoabsorption map for different thicknesses. One can see how the isoabsorption energy accurately follows the band gap values extracted from the linear fits of the Tauc plots (Figure 2a). We address the reader to the Supporting Information for detailed discussion about the analysis employed to build up the isoabsorption maps.

To gain a deeper understanding on the thickness dependence of the bandgap of atomically thin $In_2Se_3$, we performed DFT calculations, in combination with many-body techniques, as described in detail in the Supporting Information. We first investigated the electronic structure of bulk $In_2Se_3$ within DFT. As the latter tends to underestimate the electronic bandgap, we have also performed $G_0W_0$ calculations in order to get an accurate value. To access the binding energy and hence the optical gap, we included the interaction between the excited electron with the hole created in the valence band. For this we have solved the BSE starting from the $G_0W_0$ corrected DFT results.[27,28]

The evolution of the optical bandgap with the flake thickness can be modeled using a square quantum well potential of infinite height: [29,30]

$$E_{g,2D}^{opt}(d) = E_{g,bulk} - E_b + \frac{\pi^2 \hbar^2}{2d^2 \mu_{\parallel c}} \qquad (4)$$

where $E_{g,bulk}$ is the bandgap for bulk In$_2$Se$_3$ ($E_{g,bulk}$ = 1.4 eV from the $G_0W_0$ calculations), $E_b$ is the exciton binding energy ($E_b$ = 0.08 eV from the BSE calculation), $d$ is the crystal thickness and $\mu_{\parallel c}$ is the exciton reduced mass along the c-axis (perpendicular to the layers). From the $G_0W_0$ calculations we obtained an exciton reduced mass of $\mu_{\parallel c} = \left(\frac{1}{m_{e\parallel c}} + \frac{1}{m_{h\parallel c}}\right)^{-1} = 0.11\ m_e$, derived from the calculated electron ($m_{e\parallel c}$ = 0.13 $m_e$) and hole ($m_{h\parallel c}$ = 0.87 $m_e$) effective masses. The square quantum well potential, $E_{g,2D}^{opt}(d)$, as shown in Figure 2b, reproduces the dependence of $E_g^{opt}$ on the flake thickness, increasing progressively as the size of the cavity, $d$, is reduced. However, compared with the experimental data, the theoretical curve seems to be displaced in the horizontal axis towards lower values of thickness. This effect could be attributed to the surface oxidation of the In$_2$Se$_3$ crystals under ambient conditions [31], which effectively would reduce the thickness of the In$_2$Se$_3$ layer, causing an increase of the measured optical bandgap. However we do not find signatures of the presence of indium oxide in the Raman spectra measured in the crystals (see Figure S6 in the Supporting Information). On the other hand, it must be noted that the square quantum well model used here is rather simple as it does not take in account fine details of the thickness-dependent band structure of the crystal.

**Figure 3** shows a comparison of the bandgap shift observed in different atomically thin semiconductors due to the effect of quantum confinement. We observe experimentally a bandgap shift of 1.4 eV, ranging from 1.4 eV in bulk crystals to 2.8 eV in 3.1 nm thick flakes. This unusually high bandgap change, induced by quantum confinement, is among the highest observed in two-dimensional semiconductors[32,33], comparable with the one observed in atomically thin black phosphorus. Therefore, the bandgap of atomically thin $In_2Se_3$ crystals can be tuned to cover a wide region of the near ultraviolet spectrum.

In summary, we studied the effect of quantum confinement in the optical properties of atomically thin α-$In_2Se_3$ crystals. We measured the optical absorption spectra of exfoliated α-$In_2Se_3$ crystals with thicknesses ranging from 3.1 nm to 25.1 nm, observing a strong thickness-dependent shift of the optical bandgap, from 1.45 eV in the thicker flakes to 2.8 eV in the 3.1 nm thin flakes. In fact, the band gap variation observed in atomically thin $In_2Se_3$ due to the effect of quantum confinement is among the largest reported to date in 2D semiconductor materials, and is comparable to that of atomically thin black phosphorus. We performed density functional theory calculations, in combination with many-body techniques, to estimate the bandgap of the bulk $In_2Se_3$, as well as its exciton binding energy and exciton effective mass. A 2D square quantum well model allowed to reproduce the observed strong thickness dependence of the optical bandgap. In conclusion, this work shows that atomically thin *α*-$In_2Se_3$ is a very attractive and barely explored material, specially promising for applications involving tunable near-UV photodetection.

*Experimental*

*In$_2$Se$_3$ crystal fabrication*: Atomically thin In$_2$Se$_3$ flakes are prepared by mechanical exfoliation of a bulk α-In$_2$Se$_3$ crystal with Nitto Tape (Nitto Spv-224). Subsequently, we transferred these flakes to a poly-dimethylsiloxane (PDMS) substrate (Gel Film PF×4 from Gel Pak) that was selected because of its high transparency in the visible range and low interaction with atomically thin flakes [24].

*AFM measurements*: We characterized the thickness of the In$_2$Se$_3$ crystals by AFM, using a Nanotec Cervantes AFM (Nanotec Electronica) operated in contact mode. We selected contact mode AFM instead of dynamic modes of operation to avoid artifacts in the determination of the thickness. The poly-dimethylsiloxane (PDMS) substrate used during the hyperspectral imaging was inadequate for AFM measurements, as due to its viscoelastic behavior the PDMS tends to stick to the AFM tip during the scan. Therefore, it was necessary to transfer the crystals to a SiO$_2$ substrate to perform the AFM measurements, as detailed in the Supporting Information.

*Acknowledgements*
G. R-B, N.A and J.Q acknowledge financial support from MICINN/MINECO through program MAT2014-57915-R and from Comunidad Autónoma de Madrid through program S2013/MIT-3007 (MAD2D). A.C-G. acknowledges financial support from the BBVA Foundation through the fellowship "I Convocatoria de Ayudas Fundacion BBVA a Investigadores, Innovadores y Creadores Culturales", from the MINECO (Ramón y Cajal 2014 program, RYC-2014-01406) and from the MICINN (MAT2014-58399-JIN).
R. D'A and R.B. acknowledge financial support by the DYN-XC-TRANS (Grant No. FIS2013-43130-P), and NanoTHERM (Grant No. CSD2010-00044) of the Ministerio de Economía y Competitividad (MINECO), and Grupo Consolidado UPV/EHU del Gobierno Basco (Grant No. IT578-13).

[1]   K. S. Novoselov, D. Jiang, F. Schedin, T. J. Booth, V. V Khotkevich, S. V Morozov, a K. Geim, *Proc. Natl. Acad. Sci. U. S. A.* **2005**, *102*, 10451.
[2]   M. Xu, T. Liang, M. Shi, H. Chen, *Chem. Rev.* **2013**, *113*, 3766.
[3]   F. Xia, H. Wang, D. Xiao, M. Dubey, A. Ramasubramaniam, *Nat. Photonics* **2014**, *8*,

899.
[4] R. Lv, J. A. Robinson, R. E. Schaak, D. Sun, Y. Sun, T. E. Mallouk, M. Terrones, *Acc. Chem. Res.* **2015**, *48*, 56.
[5] Q. H. Wang, K. Kalantar-Zadeh, A. Kis, J. N. Coleman, M. S. Strano, *Nat. Nanotechnol.* **2012**, *7*, 699.
[6] L. Li, Y. Yu, G. J. Ye, Q. Ge, X. Ou, H. Wu, D. Feng, X. H. Chen, Y. Zhang, *Nat. Nanotechnol.* **2014**, *9*, 372.
[7] B. Radisavljevic, A. Radenovic, J. Brivio, V. Giacometti, A. Kis, *Nat Nano* **2011**, *6*, 147.
[8] Z. Yin, H. Li, H. Li, L. Jiang, Y. Shi, Y. Sun, G. Lu, Q. Zhang, X. Chen, H. Zhang, *ACS Nano* **2012**, *6*, 74.
[9] H. S. Lee, S.-W. Min, Y.-G. Chang, M. K. Park, T. Nam, H. Kim, J. H. Kim, S. Ryu, S. Im, *Nano Lett.* **2012**, *12*, 3695.
[10] W. Zhang, M.-H. Chiu, C.-H. Chen, W. Chen, L.-J. Li, A. T. S. Wee, *ACS Nano* **2014**, *8*, 8653.
[11] W. Choi, M. Y. Cho, A. Konar, J. H. Lee, G.-B. Cha, S. C. Hong, S. Kim, J. Kim, D. Jena, J. Joo, S. Kim, *Adv. Mater.* **2012**, *24*, 5832.
[12] W. Zhang, J.-K. Huang, C.-H. Chen, Y.-H. Chang, Y.-J. Cheng, L.-J. Li, *Adv. Mater.* **2013**, *25*, 3456.
[13] A. Abderrahmane, P. J. Ko, T. V Thu, S. Ishizawa, T. Takamura, A. Sandhu, *Nanotechnology* **2014**, *25*, 365202.
[14] O. Lopez-Sanchez, D. Lembke, M. Kayci, A. Radenovic, A. Kis, *Nat. Nanotechnol.* **2013**, *8*, 497.
[15] J. S. Ross, P. Klement, A. M. Jones, N. J. Ghimire, J. Yan, D. G. Mandrus, T. Taniguchi, K. Watanabe, K. Kitamura, W. Yao, D. H. Cobden, X. Xu, *Nat. Nanotechnol.* **2014**, *9*, 268.
[16] B. W. H. Baugher, H. O. H. Churchill, Y. Yang, P. Jarillo-Herrero, *Nat. Nanotechnol.* **2014**, *9*, 262.
[17] A. Pospischil, M. M. Furchi, T. Mueller, *Nat. Nanotechnol.* **2014**, *9*, 257.
[18] C.-H. Lee, G.-H. Lee, A. M. van der Zande, W. Chen, Y. Li, M. Han, X. Cui, G. Arefe, C. Nuckolls, T. F. Heinz, J. Guo, J. Hone, P. Kim, *Nat. Nanotechnol.* **2014**, *9*, 676.
[19] D. J. Groenendijk, M. Buscema, G. A. Steele, S. Michaelis de Vasconcellos, R. Bratschitsch, H. S. J. van der Zant, A. Castellanos-Gomez, *Nano Lett.* **2014**, *14*, 5846.
[20] A. Splendiani, L. Sun, Y. Zhang, T. Li, J. Kim, C.-Y. Chim, G. Galli, F. Wang, *Nano Lett.* **2010**, *10*, 1271.
[21] K. F. Mak, C. Lee, J. Hone, J. Shan, T. F. Heinz, *Phys. Rev. Lett.* **2010**, *105*, 136805.
[22] J. Yang, R. Xu, J. Pei, Y. W. Myint, F. Wang, Z. Wang, S. Zhang, Z. Yu, Y. Lu, *Light Sci. Appl.* **2015**, *4*, e312.
[23] J. Sanchezroyo, a Segura, O. Lang, C. Pettenkofer, W. Jaegermann, a Chevy, L. Roa, *Thin Solid Films* **1997**, *307*, 283.
[24] M. Buscema, G. a. Steele, H. S. J. van der Zant, A. Castellanos-Gomez, *Nano Res.* **2014**, *7*, 1.
[25] A. Castellanos-Gomez, J. Quereda, H. P. van der Meulen, N. Agraït, G. Rubio-Bollinger, *ArXiv e-prints* **2015**, 1.
[26] J. Tauc, R. Grigorovici, A. Vancu, *Phys. status solidi* **1966**, *15*, 627.
[27] M. Rohlfing, S. Louie, *Phys. Rev. B* **2000**, *62*, 4927.
[28] G. Strinati, *Phys. Rev. B* **1984**, *29*, 5718.

[29] G. W. Mudd, S. a. Svatek, T. Ren, A. Patanè, O. Makarovsky, L. Eaves, P. H. Beton, Z. D. Kovalyuk, G. V. Lashkarev, Z. R. Kudrynskyi, A. I. Dmitriev, *Adv. Mater.* **2013**, *25*, 5714.

[30] G. Bastard, E. E. Mendez, L. L. Chang, L. Esaki, *Phys. Rev. B* **1982**, *26*, 1974.

[31] C.-H. Ho, C.-H. Lin, Y.-P. Wang, Y.-C. Chen, S.-H. Chen, Y.-S. Huang, *ACS Appl. Mater. Interfaces* **2013**, *5*, 2269.

[32] W. S. Yun, S. W. Han, S. C. Hong, I. G. Kim, J. D. Lee, *Phys. Rev. B* **2012**, *85*, 033305.

[33] A. Castellanos-Gomez, *J. Phys. Chem. Lett.* **2015**, 4280.

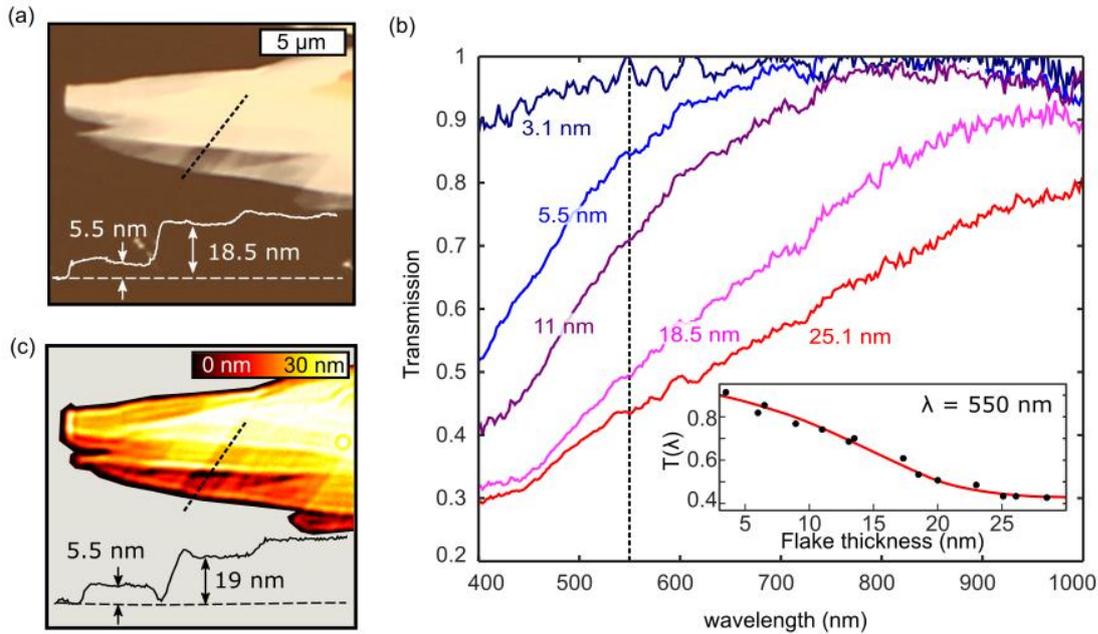

**Figure 1.** (a) Optical image in reflection mode of an $In_2Se_3$ flake with regions of different thicknesses. Inset: AFM height profile along the black dashed line. (b) Wavelength-dependent optical transmission measured in five regions of different thicknesses, from 3.1 to 25.1 nm. Inset: Optical transmission at $\lambda = 550$ nm as a function of the flake thickness. The red curve shows the interpolation of the data to a smoothed spline function. (c) Thickness colormap obtained from the hyperspectral and AFM measurements. For different illumination wavelengths we compare the transmission obtained at each image cell with the corresponding Transmission-vs-thickness curve (see inset in panel b) to estimate the flake's thickness with spatial resolution. Inset: Estimated height profile along the black dashed line, corresponding to the same region of the AFM profile shown in (a).

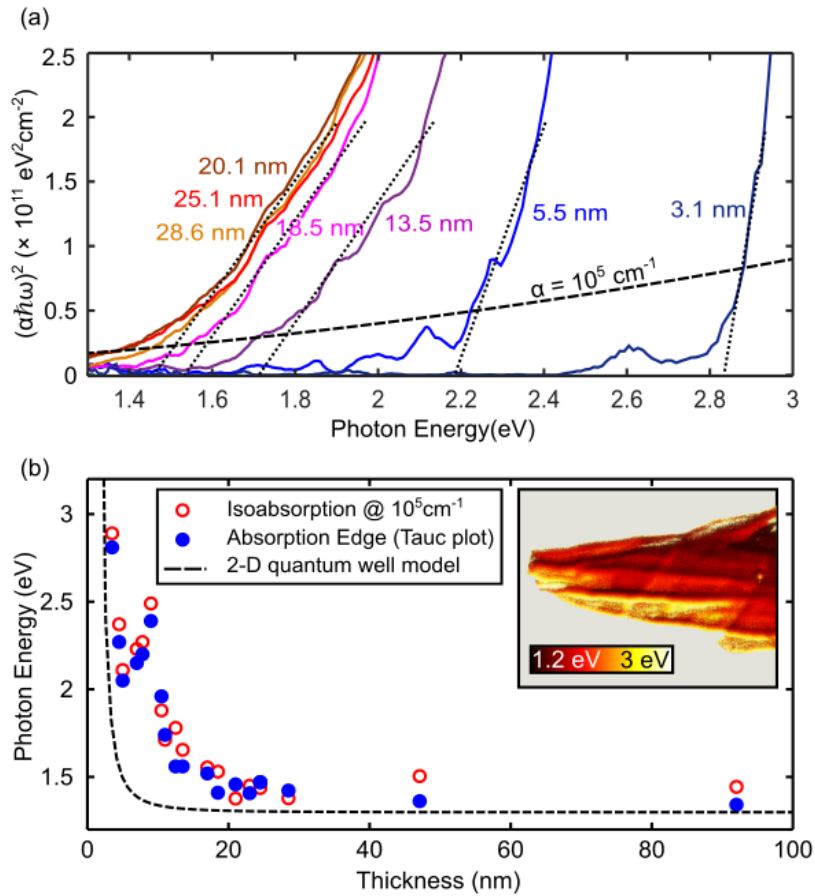

**Figure 2.** (a) Tauc plot of the optical absorption in $In_2Se_3$ flakes of 7 different thicknesses, ranging from 28.6 nm to 3.1 nm. The dotted lines show the Tauc extrapolation of the absorption edge. The energy at which the absorption coefficient takes a value of $10^5$ cm$^{-1}$ (isoabsorption energy) is shown by the black dashed curve. (b) Thickness-dependent energy of the absorption edge estimated by Tauc extrapolation (blue filled circles) and the isoabsorption at $\alpha = 10^5$ cm$^{-1}$ (red empty circles). The dashed line is a theoretical calculation of the optical gap energy using a 2-D quantum well model.

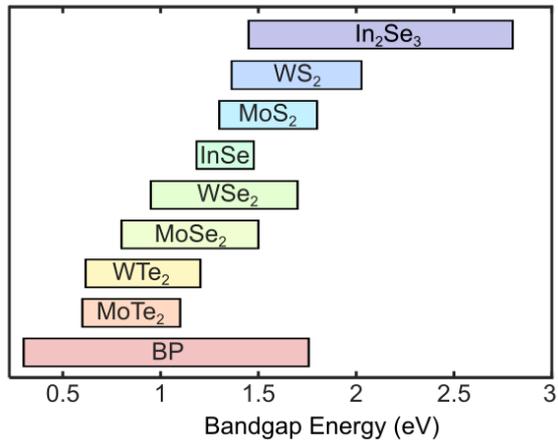

**Figure 3.** Comparison of the band gap values for different van der Waals semiconductor crystals. The horizontal bars spanning a range of energies indicate that the band gap can be tuned over that range by changing the number of layers.

# Supporting information to: Strong quantum confinement effect in the optical properties of ultrathin $In_2Se_3$

*Jorge Quereda*[1,*], *Robert Biele*[2], *Gabino Rubio-Bollinger*[1,3,4], *Nicolás Agraït*[1,5], *Roberto D'Agosta*[2,6], *and Andres Castellanos-Gomez*[5,*]

[1]*Departamento de Física de la Materia Condensada. Universidad Autónoma de Madrid, Madrid, E-28049, Spain.*
[2]*Departamento de Física de Materiales, Universidad del Pais Vasco, EHU/UPV, San Sebastian E-20018, Spain.*
[3]*Instituto de Ciencia de Materiales Nicolás Cabrera, E-28049, Madrid, Spain.*
[4]*Condensed Matter Physics Center (IFIMAC), Universidad Autónoma de Madrid, E-28049, Madrid, Spain.*
[5]*Instituto Madrileño de Estudios Avanzados en Nanociencia (IMDEA – Nanociencia), E-28049, Madrid, Spain.*
[6] *IKERBASQUE, Basque Foundation for Science, Bilbao E-48013, Spain*.
Jorge.quereda@uam.es , andres.castellanos@imdea.org

**S1. AFM measurements.**

We characterized the thickness of the $In_2Se_3$ crystals by contact-mode AFM. The poly-dimethylsiloxane (PDMS) substrate used during the hyperspectral imaging was inadequate for AFM measurements: due to its viscoelastic behavior the PDMS tends to stick to the AFM tip during the scan. Therefore, it was necessary to transfer the crystals to a $SiO_2$ substrate to perform the AFM measurements. Figure S1-a shows the same $In_2Se_3$ crystal of Figure 1a (main text) transferred onto a $SiO_2$ substrate, and Figure S1b shows an AFM topography image of the crystal.

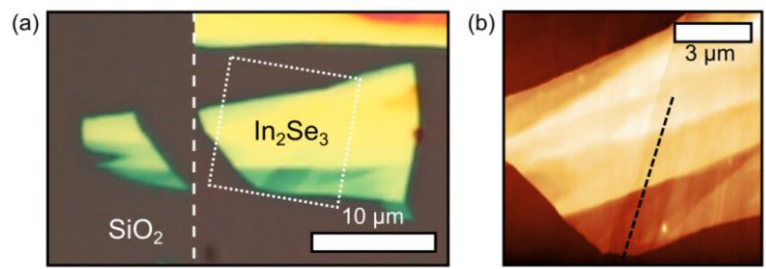

Figure S1 – (a) Transmission optical image of the same $In_2Se_3$ flake shown in Figure 1 (main text) after being transferred to a $SiO_2$ substrate. The crystal was broken in two parts during the transfer. (b) AFM topography image of the region marked by a dotted square in (a). The black dashed line corresponds to the scan shown in Figure 1a (main text).

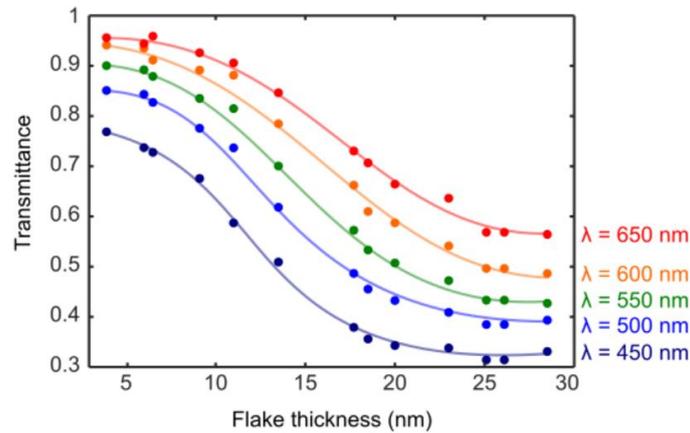

Figure S2 – Optical transmittance of $In_2Se_3$ crystals as a function of the flake thickness.

**S2. Generation of thickness maps from hyperspectral images.**

Figure S2 shows the optical transmittance of 13 different $In_2Se_3$ crystal regions as a function of their thickness. Remarkably, for a fixed illumination wavelength, the optical transmittance decreases monotonically with the flake thickness. Therefore, one can estimate the thickness of an $In_2Se_3$ crystal by quantitatively measuring the optical transmittance at a given illumination wavelength. Further, repeating this thickness estimation for each region of the optical image one can generate a two-dimensional thickness map of the $In_2Se_3$ crystal as the one shown in Figure 1 in the main text.

In order to avoid image artifacts and improve the precision of the thickness estimation, we generated 2D thickness maps for 20 different illumination wavelengths, ranging from 520 to 580 nm, and then represented their mean value. This range of wavelengths was selected because it yields a high sensitivity of the CCD camera as well as a strong thickness dependence of the optical transmittance.

**S3. Additional measurements.**

We repeated the measurements described in the main text for an additional In$_2$Se$_3$ crystal with similar results. Figure S3-a shows an optical image of the crystal under transmission illumination, as well as an AFM scan profile and Figure S3-b shows a thickness colormap of the same crystal, generated by the method explained in Section S2. The thickness map shows good agreement with the AFM measurements.

Figure S4-a shows a Tauc plot for 5 different regions of the crystal, as well as the linear fit of their absorption edges. The values of the optical bandgap estimated by the linear fit, as well as the isoabsorption energies are consistent with those shown in the main text. Figure S4-b shows an isoabsorption energy colormap of the crystal for $\alpha = 10^5 \text{cm}^{-1}$. The increase of the optical bandgap for thinnest regions is clearly observed in the colormap, consistently with the results shown in the main text.

**S4. DFT calculations**

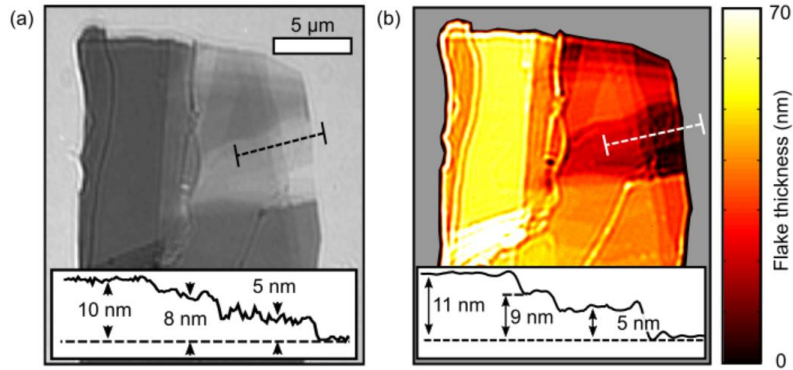

Figure S3 – (a) Transmission optical image of an $In_2Se_3$ flake with regions of different thicknesses, ranging from 3.1 nm (3 layers) to more than 90 nm. Inset: AFM scan profile along the dashed line in (a). (b) Flake thickness map generated using the method described in Section S2.

Both for In and Se, the exchange-correlation potential is described self-consistently within the generalized gradient approximation throughout the Perdew–Burke–Ernzerhof's functional (PBE). The norm-conserving Martins–Troulliers' pseudopotentials are used for both elements, additionally for In semi-core states for the valence electrons have been considered as well. We have relaxed the atomic positions with a residual force of 0.001 a.u. using the Broyden–Fletcher–Goldfarb-Shanno algorithm. The kinetic energy cutoff for the plane-wave basis set is 180 Ry, while the cutoff for the charge density is 480 Ry. The sampling of the Brillouin zone is 9 x 9 x 9 according to the Monkhorst-Pack scheme.

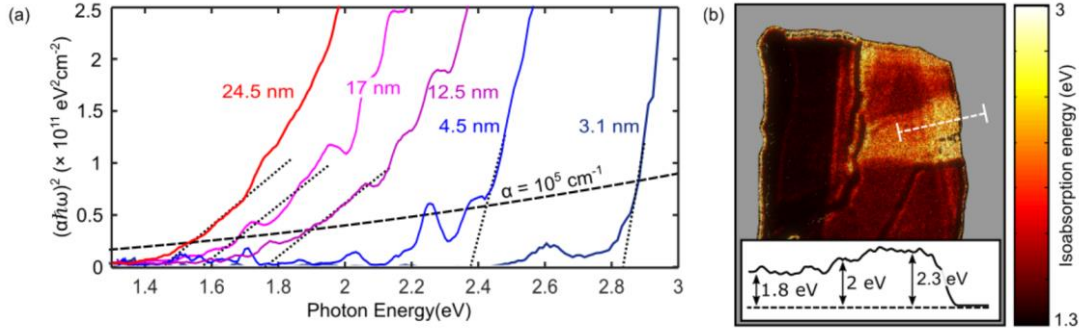

Figure S4 – (a) Tauc plot for 5 different regions of the In$_2$Se$_3$ crystal shown in Figure S1-a. The dotted lines show the Tauc extrapolation of the absorption edge. The energy at which the absorption coefficient takes a value of $10^5$ cm$^{-1}$ (isoabsorption energy) is shown by the black dashed curve. (b) Isoabsorption map of the same crystal. Inset: Isoabsorption energy profile along the region marked by a dashed line.

In order to go beyond DFT and its well known problems with the electronic bands, the electronic structure is further corrected by the non-self-consistent $G_0W_0$ approximation, where the screening is treated within the plasmon pole approximation [1, 2]. Local field effects in the screening calculations have been taken into account and we have converged the electronic quasiparticle gap within 0.02 eV. In Figure S5 the comparison between the DFT and the GW bands is shown. The highest occupied and lowest unoccupied GW band have been used to estimate the effective masses of the electron and hole in the c-direction by interpolating the bands around the $\Gamma$-point by a quadratic function. From this we have calculated the exciton reduced mass via $\mu_{\parallel c} = \left(\frac{1}{m_{e\parallel c}} + \frac{1}{m_{h\parallel c}}\right)^{-1}$.

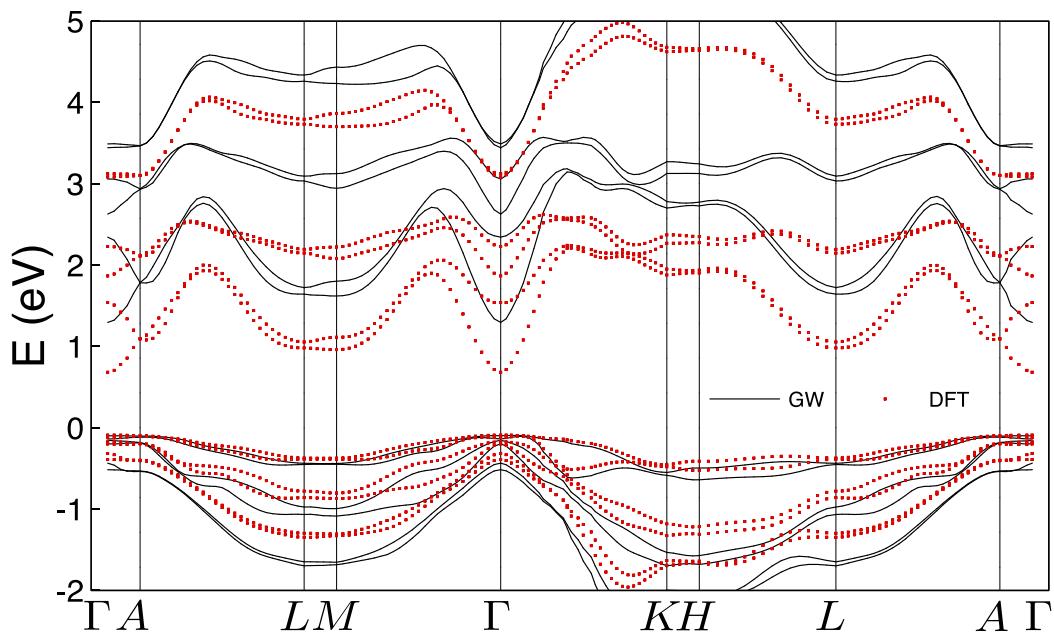

Figure S5 – Band structure for bulk In$_2$Se$_3$ within DFT and GW. While DFT underestimates the band gap, GW corrects it to 1.40 eV, which is in good agreement with the experiment. The effectives masses have been obtained by approximating the bands around the $\Gamma$-point in the c-direction by a quadratic function.

When studying the optical properties such as the absorption of light, one has to consider the interaction between the excited electron in the conduction band with the hole created in the valence band. Therefore, we have calculated the solutions of the BSE using the recursive Haydock method [3-5]. To construct the Bethe-Salpeter kernel in the static approximation we have considered 16 valence and 20 conduction bands. The position of the first peak in the optical spectrum, which corresponds to the optical gap, has been carefully converged for example with respect to the k-point sampling, the components to be summed in the exchange part and those of the screened coulomb potential of the BSE kernel. The plane-wave code Yambo is used to calculate quasiparticle corrections and optical properties with and without electron–hole effects [1], while the electronic structure calculations and structure optimisation have been performed by the DFT pseudopotentials plane-wave method as implemented in the PWSCF code of the Quantum-ESPRESSO package [2].

**S5. Raman spectra of the In$_2$Se$_3$ crystals**

Figure S6a shows four Raman spectra measured in In$_2$Se$_3$ crystals of different thicknesses, from 5 nm to 26.1 nm. As discussed in the main text, the measured spectra does not show any signature of the presence of indium oxide. Specifically, we do not find peaks at 308 cm$^{-1}$, 365 cm$^{-1}$, 504 cm$^{-1}$ and 637 cm$^{-1}$, as expected for indium oxide crystals [6, 7]. Figure S6b shows a detail of the two Raman peaks observed at ~181 cm$^{-1}$ and ~200 cm$^{-1}$, corresponding to the A$_1$ modes of α-In$_2$Se$_3$. [8]

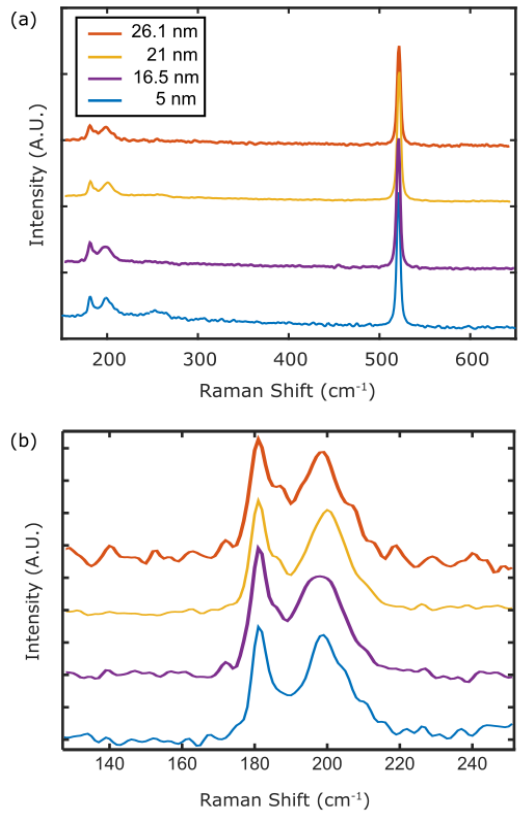

Figure S6 – (a) Raman spectra of four In$_2$Se$_3$ crystals with thicknesses ranging from 5 nm to 26.1 nm. The vertical axis have been shifted for the different signals for clarity. (b) Detail of the two peaks in the Raman spectra corresponding to the A$_1$ modes for α-In$_2$Se$_3$.

Supporting Infomation references:

1. Marini, A., et al., *Yambo: an ab initio tool for excited state calculations.* Computer Physics Communications, 2009. **180**(8): p. 1392-1403.
2. Giannozzi, P., et al., *QUANTUM ESPRESSO: a modular and open-source software project for quantum simulations of materials.* Journal of Physics: Condensed Matter, 2009. **21**(39): p. 395502.
3. Rohlfing, M. and S.G. Louie, *Electron-hole excitations and optical spectra from first principles.* Physical Review B, 2000. **62**(8): p. 4927-4944.
4. Strinati, G., *Effects of dynamical screening on resonances at inner-shell thresholds in semiconductors.* Physical Review B, 1984. **29**(10): p. 5718-5726.
5. Haydock, R., *The recursive solution of the Schrödinger equation.* Computer Physics Communications, 1980. **20**(1): p. 11-16.
6. Korotcenkov, G., et al., *Structural stability of indium oxide films deposited by spray pyrolysis during thermal annealing.* Thin Solid Films, 2005. **479**(1–2): p. 38-51.
7. Malashchonak, M.V., et al., *Photoelectrochemical and Raman characterization of $In_2O_3$ mesoporous films sensitized by CdS nanoparticles.* Beilstein journal of nanotechnology, 2013. **4**(1): p. 255-261.
8. Lewandowska, R., et al., *Raman scattering in α-$In_2Se_3$ crystals.* Materials research bulletin, 2001. **36**(15): p. 2577-2583.